\documentstyle{europhys}
\def\And{{\rm and\ }}

\def\stars{\bigskip\centerline{***}\medskip}

\newif\ifboo \boofalse

\def\Review#1{\boofalse{\it #1},}
\def\Name#1{{\sc #1},}
\def\Vol#1{\ifboo Vol. {\bf #1}\else{\bf #1}\fi}
\def\Year#1{\ifboo #1\else(#1)\fi}

\def\Page#1{\ifboo {\rm p. #1}\else{\rm #1}\fi}
\begin{document}
%
%%%   The headers.
%
%%%   These three macros are to have correct headings in your paper.
%%%   You shall omit all the arguments in the two macros `\euro{}{}{}{}'
%%%   `\Date{}' and fill in `\shorttitle{}'. 
%%%   If there is more than one author in the 
%%%   \shorttitle macro, use the macro \etal after first author's name
%%%   to obtain the correct heading.
%
\euro{xx}{x}{x-$\infty$}{1998}
\Date{19 February 1998}
\shorttitle{H. GHOSH  Complex Order parameter symmetry...}
%
%%%  The title, the Author(s) and the affiliation(s)
%
%%%   The title is set in bold (initial word only is capitalized).
%%%   Mathematical expressions and formulas within the title shall be left
%%%   in light face. Initial(s) of the first name(s) are followed by the
%%%   author(s)'s last name(s). If the authors have different affiliations,
%%%   the name must be followed by one or more \inst{number} each referring
%%%   to one of the addresses to appear in the following macro \institute.
%%%   Other items like `Present address' or `email' may be added by putting
%%%   a `\footnote' after the last \inst{number}.
%%%   Begin each address with \inst{number}; the end of an address is \\;
%%%   \\ can also be used to break a line.
%
\title{Complex order parameter symmetry and thermal conductivity}
\author{Haranath Ghosh\footnote{e-mail : hng@if.uff.br}}
\institute{
Instituto de F\'isica, Universidade Federal Fluminense,
Campus da Praia Vermelha, Av. Litor\^anea s/n,
24210 - 340, Niter\'oi, Rio de Janeiro, Brazil.}
%
%%%    The `\maketitle' macro needs the following macro:    \rec{}{}
%%%    to be left empty.
%
\rec{xx March 1998}{in final form xx xxx 1998}
%
%%%   Physics Abstracts Classification.
%
%%%   There are two macros: the first one `\pacs{}' makes the PACS 
%%%   environment,the second one `\Pacs{}{}{}' can be used for each
%%%   classification you need.
%%%   To create the subject index of the volume it is important to divide
%%%   the classification numbers into the three different arguments like
%%%   in the following examples 
%
\maketitle
%
%%%   ! Don't forget this command to format the title page of your article!
%
%%%   The Abstract
%
\begin{abstract}
Thermal behaviour of superconductors with complex order parameter
symmetry
is studied within a weak coupling theory. It is shown numerically, that the
thermal nature of the different components of complex order parametrs are
qualitatively different. Within the complex order parameter scenario, the
recent experimental observations by Krishna {\it et al.}, [Science {\bf 277},
83 (1997)] on magnetothermal conductivity and by J. Ma {\it et al.},
 [Science {\bf 267},
862 (1995)] on temperature dependent gap anisotropy for high temperature
superconductors can have natural explanation.
\end{abstract}
%
%
%%%   Main text
%
%%%   Sectioning
%
%%%   In EuroPhys there is only ``one'' level of sectioning `\section{}'.
%
An important challenge for current research is to reconcile various conflicting
results on symmetry of the superconducting order parameter in high $T_c$
cuprates by direct determination of the coexistence of other components with
$d_{x^2-y^2}$ symmetry.
Apparently, the long standing controversy concerning
$s$-wave versus $d$-wave pairing in cuprates is gradually turning in
favor of the $d_{x^2-y^2}$. For example, the precise measurements of
spontaneously generated half-integral flux quanta on bicrystal and trycrystal
films \cite{1} together with the corner SQUID expermiments \cite{2} suggest
that the superconducting gap changes sign on the Fermi surface.
Recent measurements on magnetic penetration depth \cite{3}, the nuclear
spin relaxation rate
\cite{4}, angle resolved photoemission data \cite{5}
indeed provide evidence for
 the existence of corresponding low energy excitation states.
While there exists some experimental evidence for each of these effects
\cite{6,7,8} it is by no means conclusive and an active debate continues.
 
Sun {\it et al}. \cite {9}  found a nonvanishing tunneling current along
the $c$ axis between $YBa_2Cu_3O_{6+x}$ and the conventional superconductor Pb,
which cannot exist in a pure tetragonal d-wave superconductor. It has been
 argued
by a number of authors that such data can be explained by 
considering an admixture
of $s$ wave component due to orthorhombicity in such materials \cite{10}.
A self-consistent electronic structure calculation for a $d_{x^2-y^2}$ and a
$d_{x^2+y^2}+id_{xy}$ vortex \cite{11} reveals that the scanning tunneling
spectroscopy data on vortices in $YBa_2Cu_3O_{7-\delta}$ \cite{8}
 is inconsistent with simple $d_{x^2-y^2}$ symmetry, but consistent
with $d_{x^2-y^2} + i d_{xy}$. Based on such 
recent important experimental findings,
it was suggested by Laughlin \cite{12} that there should be a tendency for
the high $T_c$ superconductors to develop a small $d_{xy}$ order parameter (OP)
on top of the usual $d_{x^2-y^2}$. Direct evidences for such suggestion are
provided by the recent outstanding experimental data on thermal conductivity
by Krishna {\it et al}., \cite{13} and theoretically, by Wheatly {\it et al,}
and most recently by
Ramakrishnan \cite{12}. We summarize below their essential findings \cite{13}
as it is found in this work that the complex order parameter symmetry has
natural explanation to this data.

A series of high resolution measurements on thermal conductivity ($\kappa$)
 in the
$Bi_2Sr_2CaCu_2O_8$ by Krishna {\it et al}. \cite{13} show that the 
$\kappa$ at low temperature becomes field-independent above a
temperature dependent threshold field $H_k(T)$.
Below $T_c$ (92 K) and in zero field a broad anomaly in $\kappa$
was observed that peaks near 65 K and an applied field ($H \parallel c$)
suppresses the anomaly.
This remarkable result
indicates a phase transition separating a low-field state where the thermal
conductivity decreases with increasing field and a high-field one where it
is insensitive to applied magnetic field. The authors argue that this phase
transition is not related to the vortex lattice because of the temperature
dependence of the field $H_k (T)$ (which is roughly proportional to $T^2$) as
well as its magnitude. Instead, they suggest a field-induced electronic
transition leading to a sudden vanishing of the quasi-particle contribution
to the heat current. Possible scenarios would be the induction of either
$id_{xy}$ or $is$ component with $d_{x^2-y^2}$ symmetry
 with application of a weak field.
Such proposition of complex order parameter symmetry in cuprate superconductors
reminds us of another important data on angle resolve phtoemission experiment
(ARPES) by J. Ma {\it et al}., \cite{14} in which a temperature
dependent gap anisotropy in
the oxygen-annealed $Bi_2Sr_2CaCuO_{8+x}$ compound was found. The measured
gaps along both high symmetry directions ($\Gamma -M$ {\it i.e}, Cu-O bond
direction in real space and $\Gamma - X$ {\it i.e}, diagonal to Cu-O bond) 
are non-zero at lower temperatures and their ratio is 
strongly temperature dependent. The experimental observation \cite{14}
is however, {\it not} reproduced by any other group.
 This observed feature cannot be
explained within a simple $d$-wave scenario and has been 
taken as a signature of a
two component order parameter, $d_{x^2-y^2}$ type close to $T_c$ and a mixture
of  both $s$ and $d$ otherwise \cite{15}. While the conductivity data
in cuprates \cite{13} indicate a finite induction of i$s$
(or i$d_{xy}$) component with
$d_{x^2-y^2}$ due to application of magnetic field, the ARPES data \cite{14}
indicates they are intrinsically so.
 
In this note, we work out the phase diagram of a superconductor with complex
order parameter symmetries such as $d_{x^2-y^2}$ + i$d_{xy}$ and
$d_{x^2-y^2}$ + i$s$ (for our purpose) comprising the amplitudes of different
components of order parameters as a function of the relative pairing
strength in different channels. It is found that the appearence of $d_{xy}$
component hardly affects the $d_{x^2-y^2}$ component 
in the case of a $d_{x^2-y^2}$ +
i$d_{xy}$ symmetry, whereas the occurence of $s$-component strongly suppresses
the $d_{x^2-y^2}$ gap
 for $d_{x^2-y^2}$ + i$s$ symmetry. These effects have been 
substantiated by calculating the temperature dependence of different components (e.g,
$\Delta_{xy}$, $\Delta_s$ and $\Delta_{d_{x^2 - y^2}}$). Interestingly enough,
the thermal growth of $d_{x^2-y^2}$ is locked at the onset of the $s$-wave
component in $d_{x^2-y^2}$ + i$s$ symmetry whereas no such strong competition
is found in case of $d_{x^2-y^2}$ +i$d_{xy}$ symmetry. 
These anomalous thermal behaviors are likely to have important impacts on
different physical properties like the specific heat, Knight shift etc.
Within this complex
order parameter scenario, we calculate electronic thermal conductivity
and the temperature dependent gap anisotropy 
($\Delta_{\Gamma-M}/\Delta_{\Gamma-X} $ Vs. T). Resemblence of the calculated
results with the observed data are remarkable.
 
Assuming the applicability of weak coupling theory to high $T_c$
cuprates, the superconducting gap equation may be written as
\begin{equation}
\Delta_k = \sum_{k^\prime} V_{kk^\prime} \frac{\Delta_{k^\prime}}
{2 E_{k^\prime}}
\tanh (\frac{\beta E_{k^\prime}}{2}),
\end{equation}
where the pairing potential $V_{kk^\prime}$ for a two component order
parameter with a separable form and the corresponding gap functions
are obtained as,
\begin{eqnarray}
V_{kk^\prime}=\sum_{j=1,}^2 V_j f_{k}^j f_{k^\prime}^j~~~ \& ~~
\Delta_k = \sum_{j=1,}^2 \Delta_j f_{k}^j
\end{eqnarray}
For a $d_{x^2-y^2}$ + i$d_{xy}$ symmetry, $V_1 \equiv  V_{d_{x^2-y^2}}$,
$V_2 \equiv  V_{d_{xy}}$, $f_{k}^1 = (\cos k_x - \cos k_y)$, $f_{k}^2=
2 \sin k_x \sin k_y$ and the corresponding component gap functions are
$\Delta_1 = \Delta_{d_{x^2-y^2}}$ and $\Delta_2 = i \Delta_{d_{xy}}$. 
Similarly, for a corresponding $d_{x^2-y^2}$ +i$s$ phase, $V_2 = V_s$, $f_{k}^2 =$
constant and $\Delta_2 =i \Delta_s$. Substituting (2) in (1) and
separating the real and imaginary parts, gap equations for different components
may be written as,
\begin{equation}
\Delta_j = \sum_{k} V_{j} \frac{\Delta_j {f_{k}^j}^2}{2 E_k}
\tanh (\frac{\beta E_{k}}{2})
\end{equation}
where the quasiparticle energy spectrum of the superconducting state
is given by $E_k =
\sqrt{(\epsilon_{k}-\mu)^2+\mid \Delta_k \mid ^2} $, 
$\epsilon_k = -2 t (\cos k_x + \cos k_y)$ being the normal state band energy. 
The coupled equations
(3) are solved numerically selfconsistently together with a number conserving equation
to fix chemical potential $(\mu$), for a given set of parameters. Then the
self-consistent values of the order parameters are used to calculate thermal
conductivity using the formula proposed by Bardeen {\it et al}. \cite{16},
long time ago.
\begin{equation}
\kappa = \sum_k \frac{(E_k v_k \cos \theta)^2}{T\Gamma}(-\frac{\partial f_{k}^0}
{\partial E_k})
\end{equation}
where $v_k \cos \theta$ is the component of group velocity parallel to
$-\nabla T$, $\Gamma$ is the relaxation rate and $f_{k}^0$ is
the Fermi distribution function.
It is important to consider the correct energy dependence of $\Gamma$ to 
include strong inelastic and impurity scatterings.
%%%%%%%%%%%%%%%%%%% Figure1%%%%%%%%%%%%%%%%%%%%%%
%\begin{figure} 
%\vfil
%\epsfxsize=4.5truein%14cm 
%\epsfysize=3.5truein%14cm 
%\centerline{\epsffile{/home/visita/hng/dwave/figure1.eps}}
%%\bigskip
%\caption{}
%\label{figure1} %fig1
%\begin{center}
%\psfig{figure=figure1.eps,height=2.5in,width=4in}
%\end{center}
%\end{figure}
%%%%%%%%%%%%%%%%%%%%%%%%%%%%%%%%%%%%%%%%%%%%%%
Phase diagrams of a $d_{x^2-y^2}$ + i$d_{xy}$ and that of a
$d_{x^2-y^2}$ +i$s$ superconductor evaluated at $T = 5 $ K
are presented in the figures 1(a) and 1(b)
respectively. The phase diagram for $d_{x^2-y^2}$ + i$d_{xy}$
superconductors  presented in Fig. 1(a) comprises the
amplitudes of $\Delta_{x^2-y^2}$, $\Delta_{xy}$ as a function of the relative
pairing strength between the two channels,
$V_{d_{x^2-y^2}}/{V_{d_{xy}}}$. When the relative pairing strength is
about 0.65 the $\Delta_{d_{xy}}$ starts developing and its appearence
affects the $\Delta_{d_{x^2-y^2}}$ component only a little, only when
$\Delta_{xy} >> \Delta_{d_{x^2-y^2}}$. A similar phase diagram for the
$d_{x^2-y^2}$ + i$s$ superconductors is presented in Fig. 1(b).
It is clear from Fig. 1(b) that with the onset of the $s$ wave
component, the $\Delta_{d_{x^2-y^2}}$ component is strongly suppressed resulting
in a pure $s$-wave phase for $V_{d_{x^2-y^2}}/{V_s} < 0.39$. This is
because of the peculiar momentum dependence of $\Delta_{d_{x^2-y^2}}(k)$ and
$\Delta_{d_{xy}}(k)$ ; the regions of the Fermi surface where
$\Delta_{d_{x^2-y^2}}(k)$ is maximum $\Delta_{d_{xy}}(k)$ is
minimum and vice versa. This is however, not the case for 
$\Delta_{d_{x^2-y^2}}(k)$
and an isotropic $\Delta_s$ symmetry. Note, while obtaining  phase 
diagrams in figures 1(a) and 1(b), the stability conditions of different
phases are also checked through free energy comparison.

%%%%%%%%%%%%%%%%%%%%Figure2%%%%%%%%%%%%%%%%%%%%%
%\begin{figure} 
%\epsfxsize=5.2truein%14cm 
%\epsfysize=3.5truein%14cm 
%\centerline{\epsffile{/home/visita/hng/dwave/figure2.eps}}
%\psfig{figure=figure2.eps,height=2.5in,width=4in}
%\bigskip
%\caption{}
%\label{figure2} %fig1
%\end{figure}
%%%%%%%%%%%%%%%%%%%%Figure2%%%%%%%%%%%%%%%%%%%%%
 The strong competition between the
$d_{x^2-y^2}$ and $s$-wave components in the $d_{x^2-y^2}$ + i$s$
phase is more distinctly 
evident in Fig. 2(b). In Fig. 2(b) temperature dependence of 
different components of the complex order parameter $d$ + i$s$ is presented
for various values of $V_{d_{x^2-y^2}}/{V_s}$ (such that both the
order parameters coexist). The curves with left arrow sign correspond to
thermal dependence of the the $s$-wave component and the rest that for the
$d$-wave component. It is shown that with the onset of $s$-wave component
the thermal growth of the $d$-wave component is arrested. This scenario is
however absent in case of a $d$ + i$d$ superconductor (cf. Fig. 2(a)). 
In Fig. 2(a),
with the decrease of the ratio $V_{d_{x^2-y^2}}/{V_{d_{xy}}}$
the $d_{xy}$ component enhances substantially but the magnitude of 
$d_{x^2-y^2}$ component
is reduced only marginally. To note, from figures 2(a) and 2(b)
 that the transition
temperature ($T_c = 84$ K) is always determined by the 
$d_{x^2-y^2}$ component irrespective
of $d$ + i$d$ or $d$ + i$s$ symmetry of the OP. Or in other words, we have
$d_{x^2-y^2}$ symmetry towards higher temperatures close to $T_c$ and an 
admixture of either
$d$ + i$d$ or $d$ + i$s$ otherwise. Furthermore, the only parameter 
that has been tuned
through out is the ratio $V_{d_{x^2-y^2}}/{V_{d_{xy}}}$
and the other cut-off parameter $\Omega_c = 0.6$
 (measured with respect to the hopping integral
$t$) is kept fixed.

According to the proposal \cite{13} (see also {12}), application of the magnetic field induces either a $is$ or a $id_{xy}$ component to the $d_{x^2-y^2}$
symmetry ({\it i.e} stronger the field more the induction of complex component) 
and thereby the magnetothermal conductivity is 
suppressed at lower temperatures with
magnetic field. In the complex order parameter scenario, we have a device to
enhance the complex component (to the $d_{x^2-y^2}$ symmetry) by reducing the
parameter $V_{d_{x^2-y^2}}/{V_{d_{xy}}}$ or
$V_{d_{x^2-y^2}}/{V_s}$.
Hence, an essentially similar physical effect can be brought in.
As a word of caution, however, it is not known whether
the magnetic field simply reduces the relative pairing strength in a
complex order parameter symmetry.
In figures 3(a) and 3(b) thermal conductivities
of a pure superconductor with complex order parameter symmtries ($d$ + i$d$ or 
$d$ + i$s$) are presented for various values of the relative pairing
strength between different channels. By pure superconductor we mean that the
impurity bound states are neglected and only 
essentially anisotropic superconducting
states which meet the case of resonance scattering is considered. 

%%%%%%%%%%%%%%%%%%%%Figure3%%%%%%%%%%%%%%%%%%%%%
%\begin{figure} 
%\epsfxsize=5.2truein%14cm 
%\epsfxsize=3.5truein%14cm 
%%\bigskip
%\caption{}
%\psfig{figure=figure3.eps,height=2.5in,width=4in}
%\label{figure3} %fig1
%\end{figure}
%%%%%%%%%%%%%%%%%%%%Figure3%%%%%%%%%%%%%%%%%%%%%

Detailed theory of thermal conductivity for
unconventional (pure) superconductors are discussed by many authors \cite{17}. 
It is worth mentioning that both the $d$ + i$d$ and $d$ + i$s$ symmetries
are parity and time reversal symmetry violating states. Also both the 
symmetries correspond to a fully gapped situation similar to an isotropic
$s$ wave superconductor. A large suppression in the normalized
thermal conductivity ($\frac{\kappa_s(T)}{\kappa_n(T)}$) at 
lower temperatures with lowering of $V_{d_{x^2-y^2}}/{V_{d_{xy}}}$ or
$V_{d_{x^2-y^2}}/{V_s}$ is seen in Fig.s 3(a, b).
Apart from a low temperature anamoly, a broad maxima is found around T = 61 K
below $T_c = 84$ K in both the cases (cf Fig. 3(a) and Fig. 3(b)).  
Also there is only a very little change in the 
conductivity at lower temperatures, 
when the ratio  $V_{d_{x^2-y^2}}/{V_{d_{xy}}}$ or
$V_{d_{x^2-y^2}}/{V_s}$ is lower enough 
(for example cf. Fig 3(a) for $V_{d_{x^2-y^2}}/V_{d_{xy}}$ = 0.446 
and 0.435). For higher values of $V_{d_{x^2-y^2}}/{V_{d_{xy}}}$ =0.555
or $V_{d_{x^2-y^2}}/{V_s}$ =0.455 the low temperature anomaly is worth
noticing. The i$d_{xy}$ or i$s$ component being small (for those 
values of $V_{d_{x^2-y^2}}/{V_{d_{xy}}}$ =0.555 or 
$V_{d_{x^2-y^2}}/{V_s}$ =0.455)
exists only upto very lower temperatures (cf. Figs. 2(a) and 2(b) also)
and hence it corresponds to a situation of a pure
$d$-wave with nodes at higher temperatures to that of a fully gapped 
$d$ + i$d$ or $d$ + i$s$ superconductor at lower temperatures. This results
in power law falling of thermal conductivity at higher temperatures to a
sudden exponential fall of the same at lower temperatures and hence explains the
lower temperature anomaly. However, such temperature dependence changes from
power law to exponential even at higher temperatures 
as the complex component i$s$ or i$d$
becomes substantial. These results are in complete agreement with the 
experimental findings \cite{13}. 
There is another extra cusp like feature seen
in case of $d$ + i$s$ superconductors ({\it e.g,} 
$V_{d_{x^2-y^2}}/{V_s}$ =0.455) at around T = 14 K. This is reflection
to the arresting of the growth of $d$ wave component with the appearence of
$s$ component seen in Fig. 2(b).
However, our figures 3(a,b) does not show 
suppression of the broad peak at around
61 K with field which is seen in experiment. 
This is because the applied field in
the experiment is temperature dependent (roughly proportional to $T^2$). 
Therefore, it turns out that the suppression in the
broad peak may be associated with
 the temperature field coupled with the magnetic
field in \cite{13}. 
So in order to see the
net effect of a pure magnetic field in thermal conductivity, it would
be suggestive
to repeat the same experiment \cite{13} with magnetic field 
which is independent 
of temperature. Finally, the temperature dependent gap anisotropy i.e, the ratio
$\Delta_{\Gamma - M}/\Delta_{\Gamma -X}$ as a function of temperature is plotted
in the inset Fig. 1(a)
 for a superconducting gap function with $d_{x^2-y^2}$ + i$d_{xy}$ symmetry.
%We estimate the gaps along the high symmetry directions from the 
%superconducting quasiparticle energy 
%spectrum $E_k = \sqrt{(\epsilon_k -\mu)^2 + \Delta_{x^2-y^2}^2(0)(\cos k_x -
%\cos k_y)^2 +  \Delta_{xy}^2 (0) \sin^2 k_x \sin^2 k_y}$.
It is seen that the temperature dependent gap anisotropy could be as high as 7
for $V_{d_{x^2-y^2}}/V_{d_{xy}}=0.446$ (at a temperature where 
the $d_{xy}$ component becomes vanishingly small). Note, in 
the $d_{x^2-y^2}$ + i$d_{xy}$ scenario, 
the 
$\Delta_{\Gamma -M} \propto \Delta_{d_{x^2-y^2}}$ and $\Delta_{\Gamma -X} 
\propto \Delta_{d_{xy}}$
($\Gamma -M \equiv (0, \pi)$ 
and $\Gamma -X\equiv (\pi/2,\pi/2)$) and therefore, the inset Fig. 1(a) can
be qualitatively estimated from Fig. 2(a). In inset Fig. 1(a), 
the large value of the
gap ratio can be obtained only when the $d_{xy}$ component is
comparable to the $d_{x^2-y^2}$ component. Similar feature is also true in case
of $d_{x^2-y^2}$ +i$s$ superconductors (not shown in figure) \cite{15}. 
While the inset Fig. 1(a) would naturally
account the experimental obseravtion by Ma {\it et al.} \cite{14}, according to 
Laughlin's conjecture \cite{12} the $d_{xy}$ component cannot be too large.

In summary, motivated by recent experimental datas concerning 
superconducting pairing
symmetry in high temperature superconductors we illustrated 
in this letter some interesting
features of the complex order parameter symmetry. 
Taking examples of $d$+i$d$ and $d$+i$s$,
it is shown that the different components of a complex OP interfer 
with each other
very differently. Thermal conductivity is calculated in the complex order 
parameter phase
and found to be in agreement with experimental observations 
based on coupled effect of temperature and magnetic field. 
The present work neither resolves pairing mechanism nor 
establishes pairing symmetry in cuprates but,
with reference to recent experimental results \cite{13,14},
the results presented in this work may have some important
 bearings to high-$T_c$
cuprate superconductors.

\stars{Financial support by Brazilian funding 
 agency FAPERJ, (project no. E-26/150.925/96-BOLSA)
and suggessions from Raimundo R dos Santos are gratefully acknowledged.}

%
%%%   Figures and Tables
%
%%%   There are two environments, one for figures and the other for 
%%%   tables.
%

%
%%%%%%%%%%%%%%%%%%%%%%%%%%%%%%%%%%%%%%%%%%%%
%\begin{figure}
%
%\vbox to 1cm{\vfill\centerline{\fbox{Here is the figure}}\vfill}
%
%\caption{Caption of  figure.}
%\label{fig1}
%\end{figure}
%%%%%%%%%%%%%%%%%%%%%%%%%%%%%%%%%%%%%55
%
%
%

\eject
\noindent {\bf Figure Captions}

\noindent {\bf Fig 1.} Phase diagram of superconductors with complex order parameter symmetries
{\bf (a)} $d_{x^2-y^2}$+i$d_{xy}$ and {\bf (b)} $d_{x^2-y^2}$+i$s$. The notations
$V_{d_{x^2-y^2}}$, $V_{d_{xy}}$, $V_s$ refer to pairing strengths
in the respective channels. The inset Fig. {\bf 1(a)} represents temperature
dependent gap anisotropy in the $d_{x^2-y^2}$+i$d_{xy}$ scenario. \\

\noindent {\bf Fig 2.} Thermal variations of different components of the
superconductong gap amplitudes in the complex order parameter {\bf (a)}
$d_{x^2-y^2}$+i$d_{xy}$ and {\bf (b)} $d_{x^2-y^2}$+i$s$ symmetry 
for various values of $V_{d_{x^2-y^2}}/V_{d_{xy}}$ and $V_{d_{x^2-y^2}}/V_s$
respectively. Strong interference between different gap components in 
$d_{x^2-y^2}$+i$s$ symmetry is worth noticing in contrast to that in 
$d_{x^2-y^2}$+i$d_{xy}$ phase. \\

\noindent {\bf Fig. 3} Normalised thermal conductivity (in arbritary units)
as a function of temperature is shown in case of {\bf (a)} 
$d_{x^2-y^2}$+i$d_{xy}$ and {\bf (b)} $d_{x^2-y^2}$+i$s$ 
pairing symmetry. (The notations $\kappa_{s,n}$ refer to thermal conductivity
in the superconducting and normal state respectively).
Loss of quasi-particle current with the enhancement
of the complex component is seen at lower temperatures. 

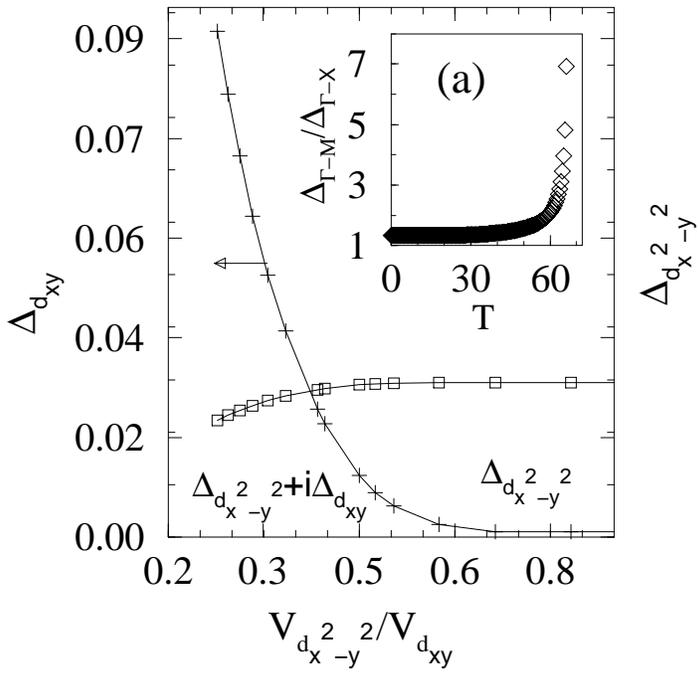
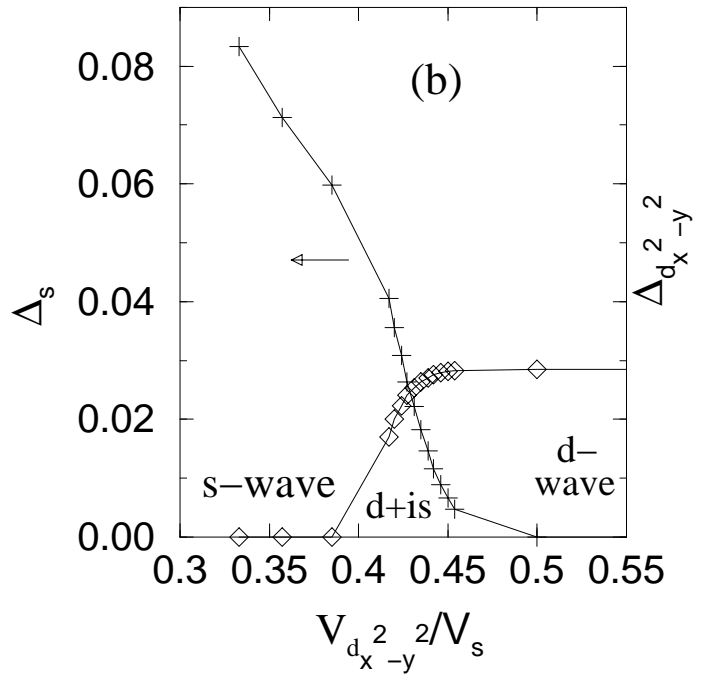
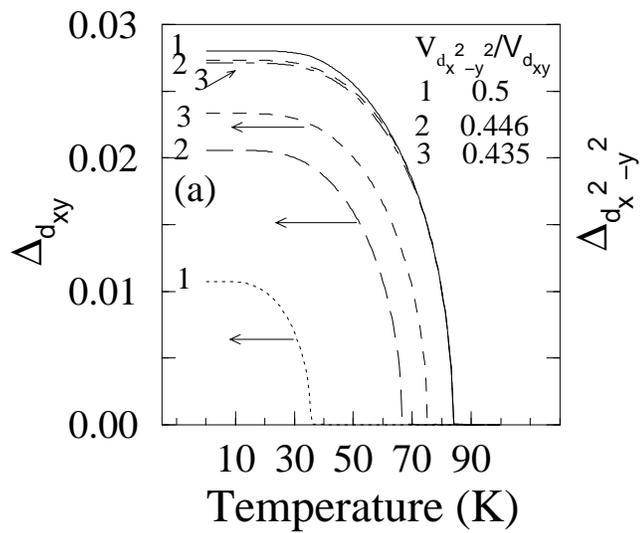
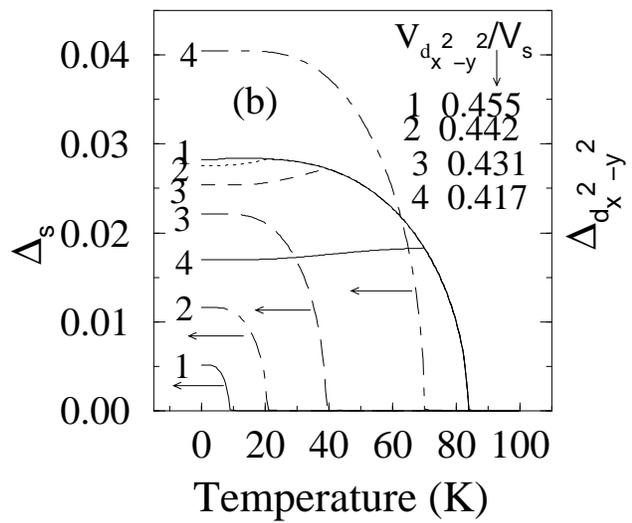
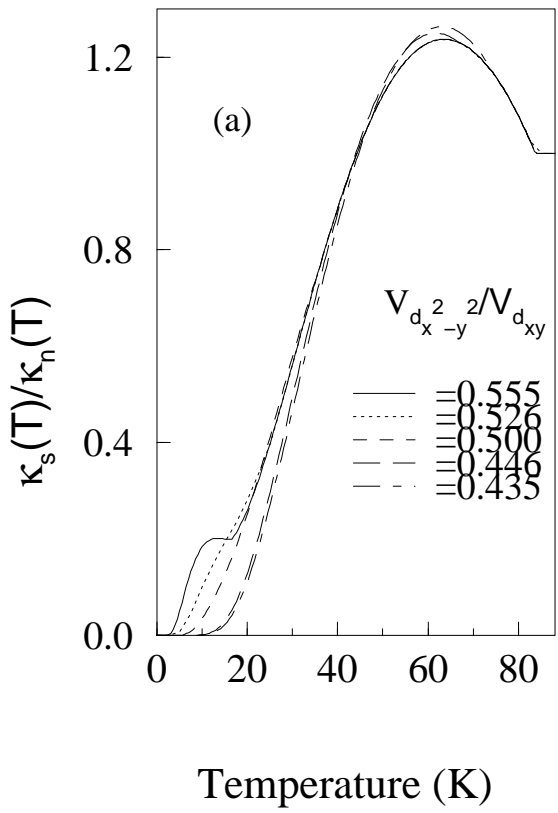
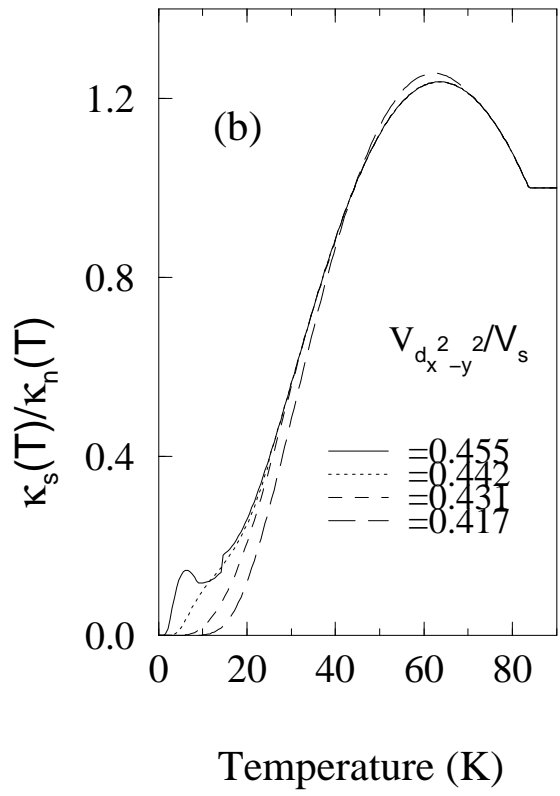
\end{document}